\documentstyle[12pt]{article}
\def\ba{\begin{eqnarray}}
\def\ea{\end{eqnarray}}
\def\bO{\mbox{$\bf \Omega$}}
\def\tbO{\tilde{\mbox{$\bf\Omega$}}}
\def\O{\Omega}
\def\tO{\tilde\Omega}
\def\o{\omega}
\def\A{{\bf A}}
\def\T{{\bf T}}
\def\R{{\bf R}}
\def\V{{\bf V}}
\def\lb{\label}
\def\be{\begin{equation}}
\def\ee{\end{equation}}

\def\D{{\bf D}}
\def\dT{{\bf dT}}
\begin{document}

\begin{titlepage}
\title{Covariant differential complexes on
quantum linear groups.
\thanks{This work is supported in part by the Russian Foundation
of Fundamental Research (grant No.93-02-3827).}}
\author{\em  A.P. Isaev\thanks{e-mail address: isaevap@theor.jinrc.dubna.su}
 and P.N.Pyatov\thanks{e-mail address: pyatov@thsun1.jinr.dubna.su}
\\
\rm Bogolubov Theoretical Laboratory, \\
JINR, Dubna, SU-101  000 Moscow, Russia}
\date{}
\maketitle

\begin{abstract}
We consider the possible covariant external
algebra structures for Cartan's 1-forms
 ($\Omega$) on $GL_{q}(N)$ and $SL_q(N)$. Our starting points are
that $\Omega$'s realize an adjoint representation of quantum group,
and all monomials of $\Omega$'s possess the unique ordering.
For the obtained external algebras we define the differential mapping
$d$ possessing the usual nilpotence condition, and
the generally deformed version of Leibniz rules. The status of the
known examples of $GL_{q}(N)$-differential calculi in
the proposed classification scheme,
and the problems of $SL_{q}(N)$-reduction are discussed.
\end{abstract}

\end{titlepage}
\newpage

\section{Introduction}
\setcounter{equation}0

Since Woronowicz have formulated
the general scheme for the
constructing of differential calculi
on quantum matrix groups \cite{Woron1},
the most publications on this theme were appealed more or less to it
(see e.g. \cite{Jurco}-\cite{Mul2}).
This scheme has the following structure:
the first order differential calculus is defined in axiomatic
way and, once it is fixed, the higher order differential
calculus can be constructed uniquely. The underlying quantum
group structure is taking into account by the bicovariance
condition.

The principal problem of the Woronowicz's approach, which have been
mentioned already in \cite{Woron1}, but still remains unsolved, is
that the scheme possesses variety of differential calculi
for each quantum group, and there is no criteria to choose
the most appropriate one.

 From the other hand, the $R$-matrix formalism (see \cite{FRT} and
references therein), initially motivated by quantum inverse
scattering method, appears to be an extremely
useful tool in dealing with quantum groups and essentially
with differential calculus on them.
So it is not surprisingly that some papers relating
Woronowicz's scheme and $R$-matrix formalism were appeared
\cite{Jurco}-\cite{Zumino}.
One may hopes, starting from the differential calculus on
quantum hyperplane and applying $R$-matrix
formulation to construct finally the most natural differential
calculus on quantum group (see \cite{Man0}-\cite{Man2}).
This program have been realized for
$GL_{q}(N)$-case in \cite{Malt}-\cite{Zum},
but when restricting to $SL_{q}(N)$
the calculus obtained reveals some unfavourable properties (see discussion
in Section 5), which force us to search for other possibilities.
So the classification of differential calculi on linear quantum groups
remains an actual problem up to now.

In the present paper we make an attempt to approach this problem
>from an opposite direction, i.e. to construct firstly the higher
order differential calculus.
Here, the key role is played by the conditions: \\
a.) Cartan's 1-forms realize the adjoint representation of $GL_{q}(N)$; \\
b.) all the higher order invariant forms, being polynomials
of the Cartan's 1-forms, can be ordered (say, lexicographically) to
unique expression.

The paper is organised as follows: all the preliminary
information and notations are collected in Section 2. In Section 3,
developing the ideas of Ref.\cite{IPy} we consider
$GL_{q}(N)$-{\sl covariant} quantum {\sl algebras} (CA). Arranging
them into two classes, the $q$-symmetrical (SCA) and
$q$-antisymmetrical (ACA) ones, we then concentrate on the
studying the homogeneous ACA's, which could be interpreted
as the external algebras of Cartan's 1-forms.
We find four one-parametric families of such algebras.
Section 4 is devoted to the construction
of differential complexes on homogeneous ACA's. In doing so
we allow the deformation of Leibniz rule and, thus, extend the
class of the permitted complexes. We conclude the paper
by the considering how the known $GL_{q}(N)$-differential
calculi is included in this scheme and discuss the problems
of the $SL_{q}(N)$-reduction.

\section{Notations}
\setcounter{equation}0

We consider
the Hopf algebra $Fun(GL_{q}(N))$ which is generated by the elements
of the $N \times N$ matrix $T=||T_{ij}||, \; i,j=1, \dots ,N$ obeying
the following relations:
\be
\lb{1}
\R\T\T' = \T\T'\R \; .
\ee
Here $\T \equiv T_{1} \equiv T \otimes  I, \;
\T' \equiv T_{2} \equiv I \otimes T, \; I$ is
$N \otimes N$ identity matrix,
$\R \equiv \hat{R}_{12} \equiv P_{12}R_{12}$,
$P_{12}$ is the permutation matrix and $R_{12}$
is $GL_{q}(N)$ $R$-matrix
\footnote{For the explicit form of $GL_{q}(N)$ $R$-matrix see
Refs.\cite{Jimbo,FRT}. }
satisfying quantum Yang-Baxter equations
and Hecke condition respectively
\ba
\R\R'\R & = & \R'\R\R' \; ,
\lb{2}        \\
\R^{2} - \lambda \R & + & {\bf 1}  =  0\; ,
\lb{3}
\ea
where $\lambda = q-q^{-1}$, $\R' \equiv
\hat{R}_{23} \equiv P_{23} R_{23}$
and ${\bf 1}$ is $N^{2} \times N^{2}$
identity matrix. In accordance with (\ref{3}),
for $q^{2} \neq -1$ the matrix $\R$ decomposes as
\be
\begin{array}{rl}
\R & = q P^{+} - q^{-1} P^{-} \; , \\
P^{\pm} & = (q + q^{-1})^{-1} \{ q^{\mp 1} {\bf 1} \pm \R \} \; ,
\end{array}
\lb{4}
\ee
where the projectors $P^{+}$ and $P^{-}$ are quantum
analogues of antisymmetrizer and symmetrizer respectively.

The comultiplication for the algebra $Fun(GL_{q}(N))$
is defined as
$\triangle T_{ij} = T_{ik} \otimes T_{kj} $ ,
and the antipode $S(.)$
\footnote{Strictly
speaking in order to define the antipodal mapping on $Fun(GL_{q}(N))$
we must add one more generator $(det_{q}T)^{-1}$ to the
initial set $\{ T_{ij} \}$ (see \cite{FRT}).}
obeys the conditions
$
S(T_{ij})T_{jl} = T_{ij} S(T_{jl}) = \delta_{il} 1 \; ,
$
so in what follows we use the
notation $T^{-1}$ instead of $S(T)$.

\section{$GL_{q}(N)$-covariant Quantum Algebras.}
\setcounter{equation}0

Consider the $N^{2}$-dimensional adjoint
$Fun(GL_{q}(N))$-comodule ${\cal A}$. We arrange
it's basic elements into $N \times N$ matrix
$A=||A_{ij}||, \; i,j=1,...,N$. The adjoint coaction is
\be
\lb{5}
A^{i}_{j} \rightarrow T^{i}_{i^{'}}
S(T)^{j^{'}}_{j} \otimes A^{i^{'}}_{j^{'}}
\equiv (TAT^{-1})^{i}_{j} \; ,
\ee
where
in the last part of the formula (\ref{5}) the standard
notation is introduced to be used below.

The comodule ${\cal A}$ is reducible, and the irreducible
subspaces in ${\cal A}$ can be extracted by use of the
so called quantum trace ($q$-trace) \cite{FRT,Reshet}
(see also \cite{Zumino,SWZum,Isaev}). In the case of $Fun(GL_{q}(N))$
it has the form:
\be
Tr_{q}A \equiv Tr({\cal D}A) \equiv
\sum_{i=1}^{N} q^{-N-1+2i} A^{i}_{i} \; , \;\;
{\cal D} \equiv diag \{ q^{-N+1},q^{-N+3}, \dots , q^{N-1} \} ,
\label{6}
\ee
and possess the following invariance property:
$$
Tr_{q}(TAT^{-1}) = Tr_{q}(A) \;,
$$
i.e. $Tr_{q}(A)$ is the scalar part of the comodule ${\cal A}$,
while the $q$-traceless part of $A$ forms the basis of
$(N^{2}-1)$-dimensional irreducible
$Fun(GL_{q}(N))$-adjoint comodule.
Let us note also the following helpful formulas:
$$
Tr_{q(2)}(\R \A \R^{-1}) = Tr_{q(2)}(\R^{-1} \A \R)
=Tr_{q}A \, I_{(1)} \; ,
$$
$$
Tr_{q(2)}\R^{\pm} = q^{\pm N} \, I_{(1)}, \;\;\;\;\;
Tr_{q}I = [N]_{q} \; ,
$$
where $\A \equiv A_{1} \equiv A \otimes I$,
$[N]_{q} = {q^{N} -q^{-N} \over q-q^{-1}}$,
and by $X_{(i)}$ we denote quantities (operators)
$X$ living (acting) in the $i$-th space.

Consider now the associative unital ${\bf C}$-algebra
$\;{\bf C}\!< A_{ij} >$
freely generated by the basic elements
of ${\cal A}$. As a vector space $\;{\bf C}\!< A_{ij} >$
naturally carries the
$Fun(GL_{q}(N))$-comodule structure. Now we introduce
$GL_{q}(N)$- {\it covariant} quantum {\it algebra} (CA) as
the factoralgebra of $\;{\bf C}\!< A_{ij} >$, possessing the
following properties \cite{IPy}:

(A) The multiplication in this algebra is defined by a set
$\{ \alpha \}$ of quadratic in $A_{ij}$ polynomial identities:
\be
\lb{8}
C^{\alpha}_{ijkl} A_{ij}A_{kl} =
C^{\alpha}_{ij} A_{ij} + C^{\alpha } .
\ee
In other words, CA is the factor algebra of $\;{\bf C}\!< A_{ij} >$
by the biideal generated by (\ref{8}).

(B) Considered as a vector space CA is a
$GL_{q}(N)$-adjoint comodule, so the coefficients
$C^{\alpha}_{ijkl}$ in (\ref{8}) are $q$-analogues
of the Clebsh-Gordon coefficients coupling two adjoint
representations, and the set of the relations (\ref{8})
is divided into several subsets corresponding to different
irreducible $Fun(GL_{q}(N))$-comodules in
${\cal A} \otimes {\cal A}$. Parameters $C^{\alpha}_{ij}$
are not equal to zero when $C^{\alpha}_{ijkl}$ couple
${\cal A} \otimes {\cal A}$
into the adjoint $GL_{q}(N)$-comodule again, while
$C^{\alpha } \neq 0$ only if
$C^{\alpha}_{ijkl} A_{ij}A_{kl} $ are the scalars.

(C) All the monomials in CA can be ordered lexicographically
due to (\ref{8}).

(D) All the nonvanishing ordered monomials in CA
are linearly independent and form the basis in CA.

Now we recall that for the classical case $(q=1)$ the
dimensions of the irreducible $Fun(GL(N))$-subcomodules
in ${\cal A} \otimes {\cal A}$
are given by Weyl formula \cite{Weyl}:
\be
dim \; {\cal A} \otimes {\cal A}  =  [(N^{2}-1) +1]^{2}  =
2 \cdot \left[ 1 \right] \oplus
(3+ \theta_{N,2}) \cdot \left[ N^{2}-1 \right] \oplus
\lb{9}
\ee
$$
2\theta_{N,2} \cdot \left [ \frac{(N^{2}-1)(N^{2}-4)}{4} \right ]
 \oplus
 \left[ \frac{N^{2}(N+3)(N-1)}{4} \right]
 \oplus \theta_{N,3} \cdot \left[ \frac{N^{2}(N+1)(N-3)}{4} \right]  ,
$$
where $ \theta_{N,M}  = \{ 1 \;for\; N>M ; \;0 \;for\; N \leq M \}$.
Thus, ${\cal A} \otimes {\cal A}$ splits into 2 scalar
subcomodules, 4 (3 for $N=2$) adjoint (traceless) subcomodules and
4 (1 for $N=2$ and 3 for $N=3$) higher-dimensional mutually
inequivalent subcomodules. In quantum case according to the
results of Ref's. \cite{LusRos} the situation generally is not
changed (the exception is for $q$ being root of unity).
Below we employ the $q$-(anti)symmetrization projectors $P^{\pm}$
and $q$-trace to extract the irreducible subcomodules
in ${\cal A} \otimes {\cal A}$, therefore supposing from
the initial that $q \neq -1$ and $Tr_{q}I = [N]_{q} \neq 0$.

First, we shall obtain the sets of quadratic in $A_{ij}$ combinations,
that correspond to the left hand side of (\ref{8}) and contain
four higher dimensional $Fun(GL_{q}(N))$-subcomodules (see (\ref{9})).
Let us start with $N^{2} \times N^{2}$ matrix $\A\R\A$
containing all the $N^{4}$
independent quadratic in $A_{ij}$ combinations and having
convenient comodule transformation properties:
\be
\lb{10}
\A\R\A \rightarrow (\T\T')\A\R\A(\T\T')^{-1} \; .
\ee
 From (\ref{1}), (\ref{4}) it follows that
$P^{\pm} \T\T' = \T\T'P^{\pm}$, hence we can split $\A\R\A$ into four
independently transforming (for $N \geq 3$) parts:
\be
\lb{11}
X^{\pm \pm} \equiv P^{\pm} \A\R\A P^{\pm} \; , \;\;\;
X^{\pm \mp} \equiv P^{\pm} \A\R\A P^{\mp} \; .
\ee
Namely the $q$-traceless (in both 1st- and 2nd- spaces) parts of
$X^{++}, \; X^{--}$ and $X^{\pm \mp}$ are the four higher dimensional
subcomodules in ${\cal A} \otimes {\cal A}$ with dimensions:
$\frac{N^{2}(N+3)(N-1)}{4}$, $\frac{N^{2}(N-3)(N+1)}{4}$
and $\frac{(N^{2}-1)(N^{2}-4)}{4}$ respectively.

Now acting on $X$'s by $Tr_{q}$-operation we obtain (for $N \neq 2$
and $q$ not beeing a root of unity) four
independent
bilinear in $A_{ij}$
combinations transforming as adjoints:
\be
\lb{12}
A^{2} \; , \;\;\; (Tr_{q}A)A \; , \;\;\; A(Tr_{q}A) \; , \;\;\;
A*A \equiv Tr_{q(2)}(\R^{-1}\A\R\A\R^{-1}) \; .
\ee
The $q$-traceless parts of these combinations correspond
to the irreducible adjoint subcomodules in ${\cal A} \otimes {\cal A}$.
Applying $Tr_{q}$ to the Eqs.(\ref{12}) once again we result in
two independent expressions
\be
\lb{13}
(Tr_{q}A)^{2} \; , \;\;\; Tr_{q}(A^{2}) \; ,
\ee
corresponding to the scalar subcomodules. We refer to the
expressions (\ref{11}), (\ref{12}) and (\ref{13}) as
higher-dimensional, adjoint and scalar terms
respectively.

As it was argued in \cite{IPy}, in order to satisfy the condition
(C) for CA, the left hand side of the relations (\ref{8}) must
contain independently either $X^{++}$ with $X^{--}$, or
$X^{+-}$ with $X^{-+}$. One can combine these pairs
into single expressions:
\be
\lb{14a}
(q+q^{-1})(X^{++} -X^{--}) = \R\A\R\A + \A\R\A\R^{-1} \; ,
\ee
\be
\lb{14b}
(q+q^{-1})(X^{-+} -X^{+-}) = \R\A\R\A - \A\R\A\R \; .
\ee
The way of the combining the quantities (\ref{11}) is not important.
We choose the concise forms (\ref{14a}), (\ref{14b}) because
in the classical limit they are
nothing but the anticommutator $[A_{2},A_{1}]_{+}$ and
commutator $[A_{2},A_{1}]_{-}$. So it is natural to call
(\ref{14a}) and (\ref{14b}) $q$-anticommutator and
$q$-commutator respectively.
In view of this all the CA's with the defining relation (\ref{8})
are classified into two types depending on whether their
defining relations contain $q$-anticommutator or $q$-commutator.
The first will be called further {\it antisymmetric} CA (ACA)
and the last -- {\it symmetric} CA (SCA).

At the moment we still fix the higher-dimensional terms in
a quadratic part of the relations (\ref{8}), but there
remains an uncertainty in the choice of the adjoint
and the scalar terms. Let us show it explicitly.
First of all we employ the simple dimensional arguments.
In order to satisfy the ordering condition (C) at a quadratic level
we must include at least
${N^{2}(N^{2}-1) \over 2}$
independent relations in (\ref{8}) (e.g. for the classical case
of $gl(N)$ it corresponds to the number of the commutators
$[A_{ij},A_{kl}]$). Since the {h-dimensional} terms (\ref{14b})
for SCA contain ${(N^{2}-1)(N^{2}-4) \over 2}$ independent
combinations, we must add to them at least $2 \cdot (N^{2}-1)$
independent combinations, i.e. two $q$-traceless adjoint terms.
Actually estimation is precise: including any other additional
adjoint or scalar terms in (\ref{8}) would result in a linear
dependence of quadratic ordered monomials and, thus, contradict
with (D). As regards the ACA's,
>from the ${N^{2}(N^{2} - 3) \over 2}$ independent combinations,
contained in h-dimensional terms (\ref{14a}), the $N^{2}$ combinations
lead to the relations
(\ref{8}) of the type: $A_{ij}^{2}=0 \; (i \neq j),
A_{ii}^{2}=\sum_{kl} f_{i}^{kl}A_{kl}A_{lk}$ (where $f^{kl}_{i}$ are
some constants), i.e. they are useless in ordering procedure.
Hence we have the deficit of the $2N^{2}$ independent quadratic
combinations  in $N \geq 3$ case (5 combinations for $N=2$) and
are forced to include in (\ref{8}) 2(1 for $N=2$) independent
q-traceless adjoint terms and a pair of scalar terms.
With this inclusion ACA's are defined by the set of
${N^{2}(N^{2}+1) \over 2}$ relations.

Thus, we have determined the number of independent adjoint and
scalar terms in symmetric and antisymmetric CA's. Note that
$q$-commutator and $q$-anticommutator itselve contain the true
number of adjoints and scalars, which is demonstrated
by the following symmetry properties:
\begin{eqnarray}
P^{\pm} \{ \R\A\R\A & - & \A\R\A\R \} P^{\pm} = 0 \; ,
\lb{15a} \\
P^{\pm} \{ \R\A\R\A & + & \A\R\A\R^{-1} \} P^{\mp} = 0 \; .
\lb{15b}
\end{eqnarray}
But there is an opportunity to change the form of quadratic adjoint
terms in the left hand side of Eq.(\ref{8}) without changing of
their number. Indeed, consider the quantities
\be
\Delta_{\pm}( U_{ad}(A) )  =  \R U_{ad}(A) \R^{\pm 1} \; \pm \; U_{ad}(A) \; ,
\lb{16a}
\ee
\be
\begin{array}{rcl}
U_{ad}(A) &=&  u^{1}(\R) \cdot \A^{2}\; + \;
(u^{2} - e)(\R)\cdot (Tr_{q}A)\A  \\
&+&  (u^{3} - e)(\R)\cdot \A (Tr_{q}A) \; + \; u^{4}(\R)\cdot (A*A) \; ,
\end{array}
\lb{16b}
\ee
where $u^{a}(\R) = u^{a}_{1} + u^{a}_{2}\R$, $a=1,2,3,4$, and
$e(\R) ={1 \over [N]_{q}}(u^{1}(\R) + q^{-N}u^{4}(\R) -{\bf 1})$.
We make the $e(\R)$-shift of the parameters $u^{2}(\R)$
and $u^{3}(\R)$
for the sake of future convenience. Expressions
$\Delta_{\pm}$ are the most general covariant combinations
which contain only adjoint and scalar (for $\Delta_{+}$) terms and
satisfy symmetry properties
\be
\lb{17}
P^{\pm}\Delta_{-} P^{\pm} = P^{\pm}\Delta_{+} P^{\mp} = 0 \; .
\ee
Therefore we may use
$\Delta_{+}$ and $\Delta_{-}$ in variating of the quadratic part
of the defining relations (\ref{8}) for ACA's and SCA's, respectively.
Note, that
in principal one could add
to the r.h.s. of (\ref{16b})
the scalar combination
$U_{sc}=h(\R)Tr_{q}(A^{2}) + g(\R)(Tr_{q}A)^{2}$, where $h$ and $g$
are arbitrary functions of $\R$.
This addition, obviously, does not affect $\Delta_{-}$.
As concerns $\Delta_{+}$, remember that defining relations for
ACA must contain a pair of independent quadratic scalars
represented in general as:
\be
\lb{17a}
Tr_{q}(A^{2}) = C_{1} Tr_{q} (A) + C_{2} \; , \;\;
(Tr_{q}A)^{2} = C_{3} Tr_{q}(A) + C_{4} \; .
\ee
Here $C_{i}$ are some constants.
Thus, even changing the form of $\Delta_{+}$, the term $U_{sc}$
can not change the content of the bilinear part of defining relations for ACA
and we will omit this term in further considerations.

 From until now we shall concentrate on studying the homogeneous
(pure quadratic) ACA's, which possess the natural $Z_{2}$-grading
and may be interpreted as an external algebras of the invariant
forms on $GL_{q}(N)$. To emphasize this step we change notations
>from $A$ to $\O$. All the other cases can be considered following
the same lines.

As we have shown, the general defining relations for homogeneous
ACA looks like
\be
 \R\O\R\O  +  \O\R\O\R^{-1} =  \Delta_{+} \; .
\lb{18a}
\ee
These relations contain 8 random parameters
$u^{a}_{i}$, $(a=1,2,3,4; \; i=1,2)$, but actually this
parametrization of the whole variety of homogeneous ACA's is redundant.
To minimize the number of parameters in (\ref{18a}), let us pass
to the new set of generators:
\be
\lb{19}
\O \rightarrow
\left \{
\begin{array}{lll}
\o  & = & Tr_{q}\O \; , \\
\tO  & = & \O - {\o \over [N]_{q}}I  \; , \;\;\; Tr_{q}\tO = 0 \; .
\end{array}
\right.
\ee
Using these new variables one can extract the first scalar
relation $\o^{2} =0$ and  (\ref{18a}) is changed
slightly to
\ba
 \R\tbO\R\tbO & + & \tbO\R\tbO\R^{-1} = \Delta_{+}(U(\tO)) \; ,
\lb{20a}  \\
\o^{2} & = & 0 \; ,
\lb{20b}
\ea
where $\Delta_{+}(U) = \R U \R  + U$ and
\be
U(\tO)  =  u^{1}(\R) \tbO^{2} + u^{2}(\R)\o\tbO
+ u^{3}(\R)\tbO\o + u^{4}(\R) (\tO * \tO) \; .
\lb{20c}
\ee
Here, as usual, $\tbO \equiv \tO_{1} \equiv \tO \otimes I$.

Applying the operations
$Tr_{q(2)}[...]$, $Tr_{q(2)}[\R^{-1}...]$, and then
$Tr_{q(1)}[...]$ to
Eq.(\ref{20a}) we
extract adjoint relations and then
obtain the second scalar relation
\be
\lb{21c}
Tr_{q}(\tO^{2}) = q^{-N}Tr_{q}(\tO * \tO) = 0 \; ,
\ee
The adjoint relations are represented in the form:
\be
\lb{21a}
v^{1}(\R)\tbO^{2} + v^{2}(\R)\o\tbO + v^{3}(\R)\tbO\o +
 v^{4}(\R)(\tO * \tO)  =  0 \; ,
\ee
where
\ba
\lb{aa21}
v^{a}(\R) =  v^{a}_{1} + v^{a}_{2} \R  & = &
x(\R)u^{a}(\R) - \delta_{a,1} q^{N} \R^{2} -
\delta_{a,4} {\bf 1} \; , \\
\lb{a21}
x(\R)=x_{0} + x_{1}\R & = &
(q^{N}+q^{-N}) + ([N]_{q} + \lambda q^{N})\R \; .
\ea
Here we arrange the pair of adjoints into a single matrix relation.
Expanding (\ref{21a}) in a power series of $\R$ one can
obtain both the adjoint relations explicitly.

Now we can reduce the number of coefficients parametrizing ACA's.
Namely, we use Eqs. (\ref{21a}) to represent some
pair of adjoint terms (\ref{12})
as linear combinations of the other two adjoints.
Let us denote $2 \times 2$ minors
of the system (\ref{21a}) as
\be
\lb{det}
\gamma^{ab} = det \left|
\begin{array}{ll}
v^{a}_{1} & v^{b}_{1} \\
v^{a}_{2} & v^{b}_{2}
\end{array}
\right| \; .
\ee
Note, if $\gamma^{34} = \gamma^{24} = 0$, then we get from
(\ref{21a}) that $\tO^{2}$ must be proportional to either
$\o\tO$, or  $\tO\o$, which contradicts with condition (D).
Hence, there are only two variants of solving (\ref{21a}) with respect to
either $\tO * \tO$ and $\tbO \o$ (if $\gamma^{34} \neq 0$), or
$\tO * \tO$ and $\o \tbO$ (if $\gamma^{24} \neq 0$).
Both choises are quite natural since, first, we exclude the
cumbersome expression $\tO * \tO$ from further considerations
and, second, we fix the order of quantities $\o$ and $\tO$
in their monomials (turning $\o$, respectively, to the left,
or to the right). In fact, as we shall see further
(see remark 3 to Theorem 1), both these
variants are equivalent and conditions $\gamma^{34} \neq 0$,
$\gamma^{24} \neq 0$ are necessary in obtaining consistent
ACA's.
So, we suppose from the initial that both $\gamma^{34}$
and $\gamma^{24}$ are not equal to zero, and choose solving
(\ref{21a}) w.r.t. $\tO * \tO$ and $\tO \o$. The result is
\be
\tO * \tO = \delta \tO^{2} + \tau \o\tO \; , \;\;
\tO\o  =  - \rho \o\tO + \sigma \tO^{2} \; ,
\lb{a23}
\ee
where
$$
\delta  = { \gamma^{13} \over \gamma^{34}} \; , \;\;
\tau = {\gamma^{23} \over \gamma^{34}} \; , \;\;
\rho = {\gamma^{24} \over \gamma^{34}} \neq 0 \; , \;\;
\sigma = - {\gamma^{14} \over \gamma^{34}} \;
$$
are namely that minimal set of parameters which we have search for.
In this parametrization the defining relations for ACA looks like
\be
 \R\tbO\R\tbO  +  \tbO\R\tbO\R^{-1} =
\bar{x}(\R) \left\{ (\delta + q^{N} \R^{2})(\R\tbO^{2}\R + \tbO^{2}) +
\tau \o(\R\tbO\R + \tbO) \right\}    \; ,
\lb{23a}
\ee
\be
\tO\o  =  - \rho \o\tO + \sigma \tO^{2} \; ,
\lb{23b}
\ee
\be
\o^{2}  =  0 \; ,
\lb{23c}
\ee
where
\ba
\lb{22}
\bar{x}(\R)  & \equiv & \{x(\R)\}^{-1} \,= \,
{1 \over [N+2]_{q}[N-2]_{q}}\left\{ - x_{2} + x_{1}\R \right\} \; , \\
\nonumber
x_{2} & \equiv & x_{0} + \lambda x_{1} \,= \,
q^{N}(q^{-2} + q^{2}) \; .
\ea
To get the relations (\ref{23a}) we solve (\ref{aa21}) with respect to
$u^a(\R)$, substitute the resulting expressions in (\ref{20c}),(\ref{20a}),
and then use (\ref{21a}) and the first of Eq's.(\ref{a23}).
The systems of relations (\ref{23a}-\ref{23c}) and
(\ref{20a}-\ref{20c}) are eqiuvalent if the matrix
$x(\R)$ is invertible (i.e. if $[N+2]_{q}[N-2]_{q} \neq 0$,
or, equivalently, $[N]_q \neq \pm [2]_q$, or $q^{2N \pm 4} \neq 1$
\footnote{C.f. with the remark in the brackets over Eq.(\ref{12}).}
). Further we shall consider this nonsingular case. The case
$N = 2$ will be treated in detail in the next Section.

Now let us discuss the symmetry properties of Eq.(\ref{23a}).
Consider the following transformation
\be
\lb{sym1}
\left\{
\begin{array}{lll}
q & \rightarrow  & q^{-1} \; ,\;\;\; hence \; ,  \;\;\;
\R_{q}  \rightarrow \R_{{1 \over q}} \; , \;\;\;
x_{q}(\cdot)  \rightarrow  x_{{1 \over q}}(\cdot) \; ; \\
\tbO & \rightarrow & \tbO' \equiv  \tO_{2} \equiv  I \otimes \tO \; .
\end{array}
\right.
\ee
Here in the lower indices we type the values of quantization parameter
for the considered quantities.
Note, that using the symmetry property of $GL_{q}(N)$
$R$-matrix:
$$
\R_{1 \over q} = P_{12} \R^{-1}_{q} P_{12}
$$
one can find that (\ref{sym1}) is a product of two symmetries:
the involution transformation of the operators ${\bf B} \rightarrow
P_{12}{\bf B}P_{12}$ and discrete symmetry
\be
\lb{sym2}
\left\{
\begin{array}{lll}
q & \rightarrow  & q^{-1} \; ,  \;\;\;
x_{q}(\cdot)  \rightarrow  x_{{1 \over q}}(\cdot) \; , \;\;\;
but \;\;\;
\R_{q}  \rightarrow \R^{-1}_{q} \; ; \\
\tbO & \longrightarrow & remains \;\;\; unchanged \; .
\end{array}
\right.
\ee
It must be stressed that the replacement $q \rightarrow q^{-1}$ doesn't
concerns the definition of $\o$, i.e. of the quantum trace. Otherwise,
we would obtain an algebra with the different covariance properties,
namely, the algebra of left-invariant (w.r.t. transitions in underlying
quantum group $GL_{q}(N)$) objects.

Now using
the identity
$$
x_q(\R_q)\R^{-1}_{q} = \frac{q^{N}\R^{2}_{q} - q^{-N}\R^{-2}_{q}}{\lambda}
$$
we deduce the following properties of the matrix function $\bar{x}(\R)$:
\ba
\nonumber
\bar{x}_{q}(\R_{q}) & = &
\R^{-2}_{q} \bar{x}_{1 \over q}(\R^{-1}_{q}) \; , \\
\nonumber
q^{N} \R^{2}_{q} \bar{x}_{q}(\R_{q}) & = &
q^{-N} \R^{-2}_{q} \bar{x}_{q}(\R_{q}) + \lambda \R^{-1}_{q} \; .
\ea
and, then, it is no hard to check that Eq. (\ref{23a}) is invariant under
the substitution (\ref{sym2}) and,
therefore, under (\ref{sym1}). In the classical limit $q=1$
this transformation reduces to the identical, but in quantum case
we get the discrete $Z_{2}$-group of symmetries of Eq. (\ref{23a}).
Namely this symmetry produce the doubling
of differential calculi on $GL_{q}(N)$ which was observed by many
authors ( see e.g. \cite{SWZum}).
\bigskip

Thus, the most general form
for the algebras, which admit an
ordering for any quadratic monomial in their generators,
is (\ref{23a})-(\ref{23c}).
The next step in finding out the consistent defining
relations for homogeneous ACA's is to consider the ordering of
cubic polynomials. Let us present the result of our considerations
in

{\bf Theorem 1:} \it
For a general values of quantization parameter $q$
there exist four
one-parametric families of homogeneous ACA's.
The defining relations for the first pair of them looks like
\be
\lb{111}
\left\{
\begin{array}{l}
\R\tbO\R\tbO + \tbO\R\tbO\R^{-1} \; = \; \kappa_{q} \left(
\tbO^{2} + \R\tbO^{2}\R \right) \; , \\
\o^{2} \; = \; 0 \; ,
\end{array}
\right.
\ee
\it and
\ba
{\;type\;\;I:}  \;\;\;\;\;\;\;\;\tO\o & = &
-\rho\o\tO\; , \;\;\;  \rho\neq 0 \; ;
\lb{112} \\
{type\;\;II:}  \;\; [ \tO , \o ]_{+} & = & \sigma\tO^{2} \; , \;\;\;
\;\;\;\;\sigma \neq 0 \; .  \lb{113}
\ea
\it Here $\kappa_{q} = { \lambda q^{N} \over [N]_{q} + \lambda q^{N} }$,
and $q \neq -1$, $[N]_{q} \neq \{ 0, \; -\lambda q^{N}, \;
-\lambda [2]_q q^{N \pm 1}, \; \pm [2]_q\}$.
For both cases the following remarkable relation holds:
\be
\lb{114}
\R\tbO^{2}\R\tbO - \tbO\R\tbO^{2}\R = 0 \; .
\ee
\it The resting pair of families can be obtained from the first one
by the  involution (\ref{sym1}) or (\ref{sym2}).

Finally in the classical limit ($q=1$) there exists one more family
of homogeneous ACA's:
\be
\lb{115}
\left\{
\begin{array}{lll}
[ \tO_{1} , \tO_{2} ]_{+} & = &
\tau' \left( P_{12} - {2 \over N} \right)
\o \left( \tO_{1} + \tO_{2} \right) \; , \\
{[} \tO , \o {]}_{+} & = & 0 \; , \\
\o^{2} & = & 0 \; .
\end{array}
\right.
\ee
\it where $\tau' = \frac{\tau N}{N^2 -4} \neq 0$
(see Eq's. (\ref{a23}),(\ref{23a})).
\rm  \\ \\
{\bf Proof:}
We shall prove the Theorem for type I and II algebras. The results
for the second pair of algebras are obviously obtained by applying
transformation (\ref{sym2}) to all the formulae below.

To check the ordering at a cubic level it is enough to consider two
monomials: $(\R\tbO)^{2}\o$ and $(\R'\R\tbO)^{3}$.
In the classical limit
these combinations become
$\tO_{2} \tO_{1} \o$ and $\tO_{3} \tO_{2} \tO_{1}$,
respectively, and for the ordinary external algebra
of invariant forms on $GL(N)$ the procedure of their
ordering looks like
$\tO_{2} \tO_{1} \o  \rightarrow \o \tO_{1} \tO_{2} $ and
$\tO_{3} \tO_{2} \tO_{1}  \rightarrow  \tO_{1} \tO_{2} \tO_{3}$.

Before establishing the quantum analog of this procedure
we have to choose the basis of "ordered" cubic monomials.
Here  the notion "ordered" is given in quotation-marks since we can't
achieve true lexicographic ordering of monomials
without loosing the compact matrix form of our considerations
and passing to cumbersome calculations in $\O_{ij}$-components.
Such an in-component calculations, based on the use of Diamond Lemma
(see \cite{Bergman}), were carried out  for the case $N=2$ in Ref's.
\cite{Man2,Malt,Mul1} and it seems doubtful that they could be
repeated for general $N$. So, we use the basis of quasi-ordered
cubic combinations, which are convenient in our matrix manipulations.
Following this way we can not prove that we have exhausted
all the possible types of ACA's, but the algebras obtained are shurely
satisfy all the conditions for ACA and
our conjecture is that the theorem 1 gives the all possible ACA's.

Let us define some new symbols:
$$
\begin{array}{lll}
(A \circ B)_{12} = \A\R{\bf B}\R^{-1}  ,&
(A \circ B)_{13} = \R'(A \circ B)_{12}\R'  ,&
(A \circ B)_{23} = \R\R'(A \circ B)_{12}\R'\R  , \\
(A)_{1} = \A  , &
(A)_{2} = \R\A\R  , &
(A)_{3} = \R'\R\A\R\R'  .
\end{array}
$$
Here, as usual, ${\bf B}  \equiv  B_{1} \equiv B \otimes I$.
The lower indices in these notations
are originated from the analogy with the classical
case $(q=1)$, where $(A \circ B)_{12} = A_{1} B_{2} , \;
(A \circ B)_{13} = A_{1} B_{3}$ etc.

We choose the following basic set of cubic matrix combinations:
\be
\lb{basis}
(\tO^{2} \circ \tO)_{ij} \;, \;\; (\tO \circ \tO^{2})_{ij} \; , \;\;
\o(\tO \circ \tO)_{ij} \; , \;\; (\tO^{3})_{i} \; , \;\; \o(\tO^{2})_{i} \; ,
\ee
where $i < j$ and $i,j = 1,2,3$. We also imply
that these basic combinations
can be multiplied from the left by any matrix function
$f(\R , \R' )$, but expressions produced from the
combinations (\ref{basis})  by
multiplication from the right are to be ordered yet.

Now in quantum case we order monomials $(\R\tbO)^{2}\o$
and $(\R'\R\tbO)^{3}$ in a following way:
\be
\lb{27}
\left\{
\begin{array}{lll}
(\R \tbO)^{2} \o & \rightarrow & - \o \tbO \R \tbO \R^{-1} + ...
\\
(\R' \R \tbO)^{3}  & \rightarrow &
- \tbO\R\R' \tbO \R{\R'}^{-1} \tbO \R^{-1}{\R'}^{-1} + ... ,
\end{array}
\right.
\ee
where by dots we denote some additional terms
which are to be expressed in terms of the
basic combinations (\ref{basis}). The point is
that such an ordering can be performed by two
different ways, depending on whether we first
permute the left pair of the generators, or the right one.
According to the condition (D) both results must be
identical, i.e. the additional terms in (\ref{27})
calculated in two ways must coincide, otherwise
the ordered cubic monomials would not be linearly
independent. Checking this condition for the combination
$(\R \tO)^{2} \o$ we get the following relation:
\be
\begin{array}{l}
\sigma \left[ (\tO \circ \tO^{2})_{12} \,+\,
(\tO \circ \tO^{2})_{21}
 \,-\,  \rho ( (\tO^{2} \circ \tO)_{12} \,+\,
(\tO^{2} \circ \tO)_{21} ) \right] \,=  \\
 \sigma \bar{x}(\R) \left[ (1-\rho) (\delta + q^{N} \R^{2})
( (\tO^{3})_{1} + (\tO^{3})_{2} )  \,+\,
\tau \o ( (\tO^{2})_{1} + (\tO^{2})_{2} )   \right] \; ,
\end{array}
\lb{28i}
\ee
where
\be
\lb{i29}
 (\tO^{2} \circ \tO)_{21} \equiv \R \tbO^{2} \R \tbO , \;\;\;
 (\tO \circ \tO^{2})_{21} \equiv \R \tbO \R \tbO^{2}
\ee
are the combinations to be expressed
in terms of the basic ones (\ref{basis}). In doing so
one can start with the relation
\be
\lb{i30}
 \R\tbO^{2} \R\tbO  -  \tbO\R\tbO^{2} \R =
\R (\tbO \Delta_{+} - \Delta_{+} \tbO)\R
\ee
which directly follows from (\ref{23a}). Here $\Delta_{+}$
is the shorthand notation for the r.h.s. of (\ref{23a}). Omitting
the straightforward but ruther tedious calculations
we present the 'ordered' expressions for $(\tO^{2} \circ \tO)_{21}$
and $(\tO \circ \tO^{2})_{21}$
in the Appendix (see (\ref{A4})-(\ref{A6})).
Substituting (\ref{A4}),(\ref{A5}) in (\ref{28i})
and considering carefully the conditions for vanishing
consequently $\o(\tO \circ \tO)_{12}$, $(\tO \circ \tO^{2})_{12}$,
$(\tO^{2} \circ \tO)_{12}$, $\o(\tO^{2})_{1,2}$
and $(\tO^{3})_{1,2}$~-terms
there we conclude that (\ref{28i}) is satisfied iff
\ba
a.) \;\; \sigma & = & 0 \; ;
\nonumber \\
b.) \;\; \sigma & \neq & 0 \;\; {\rm and} \;\; \tau = 0, \; \rho = 1 \; .
\lb{restric}
\ea

Now we  repeat these considerations for $(\R' \R \tO )^{3}$.
Performing the ordering of this expression in two different
ways we obtain the following condition
\be
\lb{31i}
\begin{array}{lllllll}
\Delta_{+} \R' \R \tbO \R^{-1} {\R'}^{-1} & - &
\R' \R \tbO \R \R' \Delta_{+} & + &
\R' \Delta_{+} \R' \R \tbO \R^{-1}  & - &  \\
\R \tbO \R \R' \Delta_{+} {\R'}^{-1} & + &
\R \R' \Delta_{+} \R' \R \tbO   & - &
\tbO \R \R' \Delta_{+}{\R'}^{-1} \R^{-1} & = & 0 \; .
\end{array}
\ee
Considering $(\tO^{3})_{1,2,3}$-terms in decomposition of (\ref{31i})
over the basic set (\ref{basis}) (here  the formulae (\ref{A4}),(\ref{A5})
are to be used) we get the condition on the parameter $\delta$
\be
\lb{delta}
\delta = - { q^{N}[N]_{q} - \lambda \over [N]_{q} + \lambda q^{N}}
\;\;\; \Leftrightarrow \;\;\;
\bar{x}(\R)(\delta + q^{N} \R^{2}) =
{ \lambda q^{N} \over [N]_{q} + \lambda q^{N}} {\bf 1} =
\kappa_{q} {\bf 1}\; ,
\ee
where $[N]_{q}  \neq - \lambda q^{N} $
is implied.
The further restrictions on the possible values of quantization
parameter $q$ are follows from the condition of invertibility
of the matrix $E(\R)$ (\ref{A6}), the inverse power of which enters
through the formulae (\ref{A4}),(\ref{A5})
all our calculations. These restrictions are
$$
\kappa_{q} \neq 1 \; , \;\; {\bf 1} + \kappa_{q} \R^{2} \not\sim P^{\pm}
\;\;\; \Leftrightarrow \;\;\;
[N]_{q} \neq 0 \; , \;\;
[N]_{q} \neq -\lambda [2] q^{N \pm 1} \; .
$$
And finally, analizing the condition (\ref{31i}) for
$\o(\tO \circ \tO)_{12,13,23}$-terms we obtain further restrictions
on parameters for case a.) (\ref{restric}):
\ba
a1.) \;\; \sigma & = & 0 \;\; {\rm and} \;\; \tau = 0 \; ;
\nonumber \\
a2.) \;\; \sigma & = & 0 \;\; {\rm and} \;\; \tau \neq 0, \; \rho = 1 \; ,
\lambda = 0 \; .
\nonumber
\ea
Checking the rest terms of Eq. (\ref{31i}) doesn't lead to the
further restrictions.

Thus, we prove the ordering conditions for cubic monomials for the
algebras (\ref{111})- (\ref{113}), (\ref{115}).
To conclude the proof of the Theorem we note that if the ordering
condition is checked  at a cubic level, then in
accordance with Manin's general remark \cite{Man1}, it automatically
follows for all the higher power monomials.
Finally, the relation (\ref{114})
follows directly from (\ref{A4}), (\ref{A5}) under the restrictions
on $\rho$, $\tau$, $\sigma$, $\delta$, that were obtained.
{\bf Q.E.D.}
\bigskip

In conclusion of the Section we make few remarks to the Theorem:
\begin{enumerate}
\item
The parameters $\sigma \neq 0$ for type II algebra
and  $\tilde{\tau} \neq 0 $  for the nonstandard
classical algebra are inessential. They can be removed from the defining
relations by simple rescalings of generators $\o$ or $\tO$.
\item
Note, that reproducing
>from the covariant relations (\ref{111})-(\ref{113}) and (\ref{115})
the explicit formulas for certain ordering prescriptions,
one may obtain some additional limitations on
the values of the parameters $\rho$, $\sigma$, $\tau$ (e.g. see below
the $N =2$ case).
\item
One can directly check the requirements (\ref{restric}) assuming
the following natural condition:
\be
\lb{y1}
Tr_{q}(\tO^{3}) \neq 0
\ee
(this is true, e.g., for the classical case $q=1$). Then,
Eqs.(\ref{a23})  lead to the relations
\be
\lb{y2}
\begin{array}{c}
[ \tO, \tO * \tO ] = \tau \left( \sigma \tO^{3} -
(1+\rho)\o \tO^{2} \right) , \\ \\
\tO^{2} \o - \rho^{2} \o \tO^{2} = \sigma (\rho - 1) \tO^{3} .
\end{array}
\ee
Applying to them the operation $Tr_{q}(.)$ and using
(\ref{y1}), (\ref{21c}) we deduce
$$
\sigma(\rho -1) = 0 = \tau \sigma
$$
which is equivalent to (\ref{restric}).
\item
Finally, we present the defining relations for homogeneous
ACA's in terms of $\O$'s (see (\ref{19}))
\be
\lb{116}
\R\bO\R\bO + \bO\R\bO\R^{-1} \; = \; \kappa_{q}
\left( \R\bO^{2}\R + \bO^{2} \right) \; + \\
\left[
\begin{array}{ll}
{\rm \;type\;I:} & {(1-\rho)(1-\kappa_{q}) \over [N]_{q} } \o
\left(\R\bO\R + \bO \right) \\
{\rm type\;II:} & {(1- \kappa_{q})\sigma \over [N]_{q} + \sigma }
\left( \R\bO^{2}\R + \bO^{2} \right)
\end{array}
\right] \; .
\ee
It should be mentioned that the condition $\rho \neq 0$
appears to be important just here, since the relations (\ref{116})
for type I algebra contain both scalar terms only under this
restriction.

\section{ Differential Complexes of Invariant Forms.}
\setcounter{equation}0

As we argued before, among the algebras presented in Theorem 1
there exists the true algebra (or maybe the set of such algebras)
of invariant differential
forms on $GL_{q}(N)$. To make the connection
with the differential calculi on quantum groups more clear we shall
supply the homogeneous ACA's listed in the Theorem 1 with a grade-1
nilpotent operator  $d$ of
external derivation. The {definition} of $d$ must respects
the covariance properties (\ref{5}) of Cartan 1-forms, i.e.
$d$ must commute with the adjoint $GL_{q}(N)$-coaction on $\O$.
Hence, the following general anzats is allowed:
\be
\lb{26}
\left \{
\begin{array}{lll}
d\cdot \tilde{\Omega} & = &
x\tilde{\Omega}^{2} + y \o \tilde{\Omega} - z \tilde{\Omega}\cdot d \\
d\cdot \o & = & - t \o\cdot d \; .
\end{array}
\right.
\ee
Here $x, \; y, \; z$ and $t$ are some parameters to be fixed
below.
We stress that the last term in the right hand side of (\ref{26})
defines the deformed version of Leibniz rules for differential forms.
The ordinary Leibniz rules are restored under the limit
$z=t=1$. Note, that for the differential calculi on the
quantum hyperplane the deformed version of the Leibniz rules
have been considered in \cite{Kul}.

Now it is straightforward to obtain

{\bf Theorem 2:} \it Under the restrictions of Theorem 1 there exists two
distinct covariant differential complexes for type I algebras,
defined by
\be
\lb{27a}
\qquad\; { type IA:} \; \left\{
\begin{array}{lll}
d\cdot \tilde{\Omega} & = & \tilde{\Omega}^{2} -
\tilde{\Omega}\cdot d ,\\
d\cdot \o & = & - \rho \o\cdot d ;
\end{array}
\right.
\ee
\be
\lb{27b}
\;\;\;\; \qquad { type IB:} \; \left\{
\begin{array}{lll}
d\cdot \tilde{\Omega} & = & \o
\tilde{\Omega} - z \tilde{\Omega}\cdot d , \\
d\cdot \o & = & - \o\cdot d .
\end{array}
\right.
\ee
\it The differential complexes for type II and the nonstandard classical
algebras are defined uniquely:
\be
\lb{27c}
\;\;\;\;\qquad { type II:} \;
\left\{
\begin{array}{lll}
d\cdot \tilde{\Omega} & = &
\tilde{\Omega}^{2} - \tilde{\Omega}\cdot d , \\
d\cdot \o & = & - \o\cdot d ;
\end{array}
\right.
\ee
\be
\lb{27d}
\begin{array}{c}
nonstandard \\ classical \; case:
\end{array}
\left\{
\begin{array}{lll}
d\cdot \tilde{\Omega} & = & \o
\tilde{\Omega} - \tilde{\Omega}\cdot d, \\
d\cdot \o & = & - \o\cdot d .
\end{array}
\right.
\ee
\it Here all the inessential parameters are removed by $\o$- and
$\tilde{\Omega}$-rescalings. \rm

{\bf Proof:} These restrictions are easily obtained by
demanding $d^{2}=0$ and checking the compatibility of
anzats (\ref{26}) with the algebraic relations
(\ref{111})-(\ref{113}).
We would like only to mention that the relation (\ref{114}) plays
an important role
when elaborating the type I and II cases. {\bf Q.E.D.}

Let us discuss which of the differential complexes
listed in Theorem 2 can be treated as $q$-deformations of
the complex of right-invariant forms on $GL(N)$.
Comparing the formulae (\ref{111}-\ref{113}) and (\ref{27a}-\ref{27c})
with the conventional classical relations:
\be
\lb{classic}
{[} \tO, \o {]}_+ = {[} \tO_1, \tO_2 {]}_+ = 0 \; , \;\;\;
d \cdot \tO = \tO^2 - \tO \cdot d \; , \;\; d \cdot \o = - \o \cdot d \; ,
\ee
we conclude, that there are two different possibilities to deform
the complex of $GL(N)$-invariant differential forms.
The first is realized by the type IA differential complexes
with the additional restriction on parameter
$lim_{q \to 1} \rho = 1$. Note, that in this case
the Leibniz rules are deformed under quantization (for $\rho \neq 1$).
The second possibility is realized by the type II differential complexes
with $lim_{q \to 1} \sigma = 0$. Here the Leibniz
rules take their conventional form. We would like to mention that
all the other types of differential complexes listed in Theorem 2
also may be interested as an examples of 'exotical'
differential complexes on $GL(N)$ and $GL_q(N)$, but this
subject lies beyond the scope of the present paper.
\footnote{For $N=2$ such an 'exotical' complexes have been
considered in \cite{Mul1}.}

Now let us treat the $SL_q(N)$-case. The $q$-traceless generators
$\tO_{ij}$ can naturally be identified with the $(N^2 -1)$-dimensional
basis of right-invariant 1-forms on $SL_q(N)$. These generators form
the closed algebra under external multiplication given in the Theorem 1
(see (\ref{111})), and, remarkably, the algebra of these generators
doesn't contain any random parameters. As the Theorem 2 states,
the action of
external derivative on these generators can be only
defined like in the classical case: $d\cdot \tO = \tO^2 - \tO\cdot d$
(see (\ref{27a}), or (\ref{27c})).
So, we conclude that the
complex of $SL(N)$-invariant differential forms possess
the unique $q$-deformation.

In the classical Lee-group theory the differential complex
of invariant forms serves as the suitable basis in the whole
de-Rahm complex of all the differential forms on the group
manifold. So, in order to get the full differential calculi
on the linear quantum groups we have to supply the algebras
obtained with the suitable cross-multiplication rules for
$T_{ij}$ and $\O_{ij}$, and to define additionaly the action of external
derivative on $T_{ij}$. Note, that
in Woronowicz's sheme \cite{Woron1} this questions
are to be solved in the first place, when constructing the first order
differential calculus. Not tempting to solve the problem
in general we present here one example of such construction,
and establish the correspondence between our homogeneos ACA's
and the existing examples
of $GL_q(N)$-bicovariant differential calculi.

For the matrix group $GL_q$ of a general rank $N$
two versions of differential calclus have been considered.
They were obtained first in the local coordinate representation,
where the differential algebra is generated
by the coordinate functions $T_{ij}$, their differentials $dT_{ij}$,
and derivations $D_{ij}$ $\left( {\rm means }\;
\frac{\partial}{ \partial T_{ji}} \right)$.
We present here the full set of relations between such generators:
\be
\lb{28}
\begin{array}{rcl}
\R\,\T\,\T' & = & \T\,\T'\,\R ,  \\
\R\,\dT\,\dT' & = & - \dT\,\dT'\,\R^{-1} , \\
\R\,\dT\,\T' & = & \T\,\dT'\,\R^{-1} ,
\end{array}
\ee
\be
\lb{28a}
\begin{array}{rcl}
\R\,\D'\,\D & = & \D'\,\D\,\R ,  \\
\D\,\R\,\T & = & {\bf 1} \,+\, \T'\,\R^{-1}\,\D', \\
\D\,\R\,\dT & = & \dT'\,\R^{-1}\,\D'.
\end{array}
\ee
Here, as usual, $\D = D\otimes I$,
$\D' =  I\otimes D$,
$\dT = dT\otimes I$,
$\dT' =  I\otimes dT $.
This algebra is checked to possess unique ordering
for any quadratic and cubic monomials. The relations
(\ref{28}) were obtained in \cite{Malt} and in
$R$-matrix formulation in \cite{Sud,Schir}. The first
two of relations (\ref{28a}) were appeared in \cite{IPy,Zum}.
Note, that the algebra (\ref{28}),(\ref{28a}) implies
the commutativity of derivations $D$ and external derivative $d$.
The defining relations for the second version of
differential calculus can be obtained
>from (\ref{28}), (\ref{28a}) by the symmetry transformation
$\R\leftrightarrow\R^{-1}$ of the type (\ref{sym2}).

The right-invariant 1-forms and vector fields are
then constructed as
\be
\lb{29}
\begin{array}{rcl}
\Omega = dT \cdot T^{-1} & , & V = T\cdot D,
\end{array}
\ee
and they possess the following algebra
\ba
\R\,\bO\,\R\,\T &=& \T\,\bO'\; , \;\;\;\;
\R\,\V\,\R\,\T\, = \,\T\,\V'\, + \,\R\,\T \; ,
\lb{30a}
\\
\lb{30}
\R\,\bO\,\R\,\bO &=& - \,\bO\,\R\,\bO\,\R^{-1} \; ,
\\
\lb{31}
\R\,\V\,\R\,\V &=& \V\,\R\,\V\,\R\, + \,\R\,\V\, - \,\V\,\R \; ,
\\
\lb{a30}
\R\,\bO\,\R\,\V &=& \V \,\R\,\bO\,\R^{-1}\, + \,\R \bO \; .
\ea
Here Eq's.(\ref{30}) are the commonly used commutation
relations for $GL_q(N)$-invariant differential forms
(see \cite{Schir}-\cite{Zum}).
Comparing (\ref{30}) with (\ref{116}) we see that $\O$'s (\ref{29})
realize the special case of type II external algebra with
$\sigma = - \kappa_q [N]_q$. Eq's.(\ref{31}) is the well known
commutation relations for $GL_q(N)$-invariant vector fields
\cite{Jurco}-\cite{Zumino}, but in a slightly different notations.
To obtain this relations in the conventional form we have to
pass to the new basis of generators
${\bf Y} = 1- \lambda \V $.
In this basis Eq's.(\ref{a30}),(\ref{31}) looks like
\ba
\lb{31a}
\R\,{\bf Y}\,\R\,{\bf Y} &=& {\bf Y}\,\R\,{\bf Y}\,\R \; ,
\\
\lb{a30a}
\R\,\bO\,\R\, {\bf Y} &=& {\bf Y} \,\R\,\bO\,\R^{-1} \; .
\ea
Note, that our commutation relations of $V$'s with $\O$'s or $Y$'s
(\ref{a30}), (\ref{a30a})
are different from that presented in \cite{SWZum,Zum}
for invariant 1-forms and Lie derivatives.

The operator of external derivation in (\ref{28}),(\ref{28a})
admits the following explicit representation
\be
\lb{d}
d = Tr_{q}(\O V Y^{-1}) \equiv Tr_{q}(dT D (1-\lambda V)^{-1}),
\ee
which surprisingly differs from expected formula $Tr_q(dT D) = Tr_q(\O V)$.
The operator (\ref{d}) satisfy the nilpotence condition and
the ordinary Leibniz rules. The form of relation (\ref{d})
suggest us an idea of changing the definition (\ref{29})
of invariant vector fields. Indeed, consider the new set of generators
$U_{ij}$ wich is obtained from the old $V$'s by the nonlinear
invertible transformation:
\be
\lb{U}
U = \frac{V}{I - \lambda V} \; , \;\;\;\;
V = \frac{U}{I + \lambda U} \; .
\ee
With a little algebra one can check that the commutation relations
(\ref{30a}),(\ref{31}), (\ref{a30}) can be concisely rewritten in terms of
$U$'s
\ba
\R^{-1}\,{\bf U}\,\R^{-1}\,\T &=& \T\,{\bf U}' \, + \,\R^{-1}\,\T \; ,
\\
\R^{-1}\,{\bf U}\,\R^{-1}\,{\bf U} &=& {\bf U}\,\R^{-1}\,{\bf U}\,
\R^{-1}\, + \,\R^{-1}\,{\bf U}\, - \,{\bf U}\,\R^{-1} \; ,
\\
\R\,\bO\,\R^{-1}\,{\bf U} &=& {\bf U}\,\R\,\bO\,\R\, + \,\R\,\bO \; .
\ea
Now, if we consider $U_{ij}$ instead of $V_{ij}$
as invariant vector fields on $GL_q(N)$, then
the formula for external derivative takes it standard form:
$d = Tr_q(\O U)$.

Finally, let us consider the simplest case of $GL_{q}(2)$-covariant
differential calculus in more details. Note, that while
the proof of Theorem I does not work for $N=2$ the resulting
formulae are applicable to this case as well. This can be
directly checked by using only the general properties of
$\R$-matrix, namely Yang-Baxter equation, Hecke condition, and the
$q$-trace formula. The fail of the general proof of Theorem I
is due to the different structure of $Ad^{\otimes2}$
decomposition in case $N=2$ and is not crucial.

Denote the components of matrix $\Omega$ as
$ \left(
\begin{array}{cc}
\theta_{1} & \theta_{2} \\
\theta_{3} & \theta_{4}
\end{array}
\right)$.
Then from the covariant expressions (\ref{111}-\ref{113})
the following explicit ordering prescriptions can be extracted:
\be
\lb{33}
\begin{array}{llcl}
type\; I: &
\theta^{2}_{2} & = & \theta^{2}_{3}  \> = \> 0 \; ,\qquad  \qquad \>
\theta_{3}\theta_{2} \> = \> - \theta_{2}\theta_{3} \; ,
\\  \\
& \theta_{1}^{2} & = &
\frac{\mbox{$1$}}{\mbox{$q^{-2} + \rho$}} \left\{
q\lambda\rho \theta_{2} \theta_{3}
\,+ \,(\rho -1) \theta_{1} \theta_{4} \right\} \; ,
\\  \\
& \theta_{4}^{2} & = &
\frac{\mbox{$1$}}{\mbox{$q^{-2} + \rho$}} \left\{
q^{-3}\lambda \theta_{2} \theta_{3}
\, - \, q^{-2}(\rho - 1)\theta_{1} \theta_{4} \right\}  \; ,
\\  \\
& \theta_{4}\theta_{1} & = & -
\frac{\mbox{$1$}}{\mbox{$q^{-2} + \rho$}} \left\{
(1 + q^{-2}\rho) \theta_{1}\theta_{4}
\, + \, q^{-1}\lambda (1 + \rho)
\theta_{2} \theta_{3} \right\} \; ,
\\  \\
& \theta_{3}\theta_{1} & = &
\frac{\mbox{$1$}}{\mbox{$ 1 + \rho$}}
\left\{  - \rho (1 + q^{2})\theta_{1} \theta_{3} \, + \, (\rho - q^{2})
\theta_{3} \theta_{4} \right\} \; ,
\\  \\
& \theta_{4} \theta_{3} & = &
\frac{\mbox{$1$}}{\mbox{$ 1 + \rho$}}
\left\{ - (1 + q^{-2})
\theta_{3} \theta_{4} \,+ \,(1 - q^{-2}\rho)
\theta_{1} \theta_{3} \right\} \; ,
\\  \\
& \theta_{2} \theta_{1} & = &
\frac{\mbox{$1$}}{\mbox{$q^{-2} + q^{2}\rho$}} \left\{
- (1 + q^{-2})\rho \theta_{1} \theta_{2}
\,+\, (q^{2}\rho - 1) \theta_{2} \theta_{4} \right\} \; ,
\\   \\
& \theta_{4} \theta_{2} & = &
\frac{\mbox{$1$}}{\mbox{$q^{-2} + q^{2}\rho$}} \left\{
- (1 + q^{2})\theta_{2} \theta_{4}
\,+\, (q^{-2} - \rho) \theta_{1} \theta_{2} \right\} \; ;
\end{array}
\ee
$$
\begin{array}{llcl}
type\; II:\;\> &
\theta^{2}_{2} & = & \theta^{2}_{3}  \> = \> 0 \; ,\qquad  \qquad \>\>\;\;\>
\theta_{3}\theta_{2} \>\> = \>\> - \theta_{2}\theta_{3} \; ,
\\ \\
& \theta_{1}^{2} & = &
\frac{\mbox{$\mu$}}{\mbox{$1 - \mu$}} \theta_{2} \theta_{3}
 \; , \qquad  \qquad
\theta_{4}^{2} \>\>\>\>\>\> = \>\>
\frac{\mbox{$1 - q^{-2} -\mu$}}{\mbox{$1 - \mu$}}
\theta_{2} \theta_{3}  \; ,
\\ \\
& \theta_{4}\theta_{1} & = & - \theta_{1}\theta_{4}
\, + \, \lambda
\frac{\mbox{$(q + q^{-1})\mu - q$}}{\mbox{$1 - \mu$}}
\theta_{2} \theta_{3} \; ,
\\  \\
& \theta_{3}\theta_{1} & = & -
\frac{\mbox{$1$}}{\mbox{$ 1 - \mu(1+q^{-2})$}}
\left\{ (1 - \mu)\theta_{1} \theta_{3} \,+ \,
\frac{\mbox{$\mu$}}{\mbox{$q^{2}$}}
\theta_{3} \theta_{4} \right\} \; ,
\\  \\
& \theta_{4} \theta_{3} & = & -
\frac{\mbox{$1$}}{\mbox{$ 1 - \mu(1+q^{-2})$}}
\left\{
\frac{\mbox{$1 - \mu$}}{\mbox{$q^{2}$}}
\theta_{3} \theta_{4} \,+\, (\mu - 1 + q^{-2})
\theta_{1} \theta_{3} \right\} \; ,
\end{array}
$$
\be
\lb{32}
\begin{array}{lcl}
\theta_{2} \theta_{1} & = & - (1 - \mu)\theta_{1} \theta_{2}
\,+ \, \mu \theta_{2} \theta_{4} \; ,
\\  \\
\theta_{4} \theta_{2} & = & - q^{2}(1 - \mu)\theta_{2} \theta_{4}
\,+ \, q^{2} (\mu  - 1 + q^{-2})\theta_{1} \theta_{2} \; .
\end{array}
\ee
Here we use parameter $\mu =
\frac{\mbox{$\sigma + [N]_{q} \kappa_{q}$}}{\mbox{$\sigma
+ [N]_{q}$}}\mid_{N=2}$ instead of $\sigma$ for sake of convenience.
In this notation the case (\ref{30}) corresponds to $\mu=0$.
An obvious restrictions $\mu\neq\left\{1,(1+q^{-2})^{-1}\right\}$
and $\rho\neq\left\{-1,-q^{-2},-q^{-4}\right\}$ arise when passing
>from covariant relations to formulation in components
(see remark 2 to the Theorem I).

Let us compare these results with that presented for $GL_{p,q}(2)$
case in Ref.~\cite{Mul1}.First, we note that
by assumption the left-invariant 1-forms in \cite{Mul1}
admit the decomposition $\Omega=T^{-1}\cdot dT$, and the
external derivative satisfies the undeformed version of Leibniz rools.
Hence, the formula $d\cdot\Omega= - \Omega^{2} - \Omega\cdot d$
is postulated. Moreover, the relation $d\Omega \sim \bigl[\omega,
\Omega\bigr]_{+}$ is also implied. Hence, the differential calculi
obtained in \cite{Mul1} must be of the second type.
Indeed the relations (8.5) of \cite{Mul1} can be transformed
into the form (\ref{32}) if we note that due to the conditions
(6.25) \cite{Mul1} for parameter $\bf N$ (see (8.1-3) \cite{Mul1} )
the following quadratic relation is satisfied:
\be
\lb{34}
(1+r^{2})\left( 2+ \mbox{\bf N}\left( 1+r-(1+r^{2})s \right) \right)
 = \left( 1+r+r\mbox{\bf N} \right)^{2} \; .
\ee
Here $r = pq$ (see Eq. (8.5) \cite{Mul1}) is the only combination of
deformation parameters that enters the external algebra of invariant
forms. This is not surprisingly since $GL_{p,q}(2)$ \R-matrix,
when suitably normalized, satisfies Hecke relation
\be
\lb{35}
\R^{2} = \mbox{\bf 1} + \bigl(r^{\frac{1}{2}} - r^{-\frac{1}{2}}
\bigr)\R \; .
\ee
Hence, we expect that parameter $r^{\frac{1}{2}}$ of
\cite{Mul1}
corresponds to ours $q^{-1}$ ( the inverse power here is due to
the substitution $q \leftrightarrow q^{-1}$ that should be done
to pass from the right-invriant forms of Eq. (\ref{32}) to
the left-invariant ones).

Variable $s$ of (\ref{34}) parametrizes the different
external algebras in
\cite{Mul1} and it should be corresponded
to ours $\mu$. Actually, using (\ref{34}) it is strightforward
to check that Eqs. (8.5) of
\cite{Mul1} are equivalent to
(\ref{32}) with the following substitutions to be made:
$$
\begin{array}{ccc}
r \leftrightarrow q^{-2} & , &
\frac{\mbox{$r^{-1} - 1 - r\mbox{\bf N}$}}{\mbox{$r + r^{-1}$}}
 \leftrightarrow \mu \; .
\end{array}
$$

Sumarizing all the above, we conclude that our
type II differential complexes
>from one hand generalize
for arbitrary $N$ the $N=2$ the formulae given in \cite{Mul1}
and from another hand include the bicovariant calculi considered
in \cite{Malt}-\cite{Zum}.

\section{Conclusion}
\setcounter{equation}0

Here we make some comments on constructing the differential
calculi for type II complexes (\ref{111}),(\ref{113}),(\ref{27c}),
and discuss briefly the problem of $SL_q(N)$-reduction of the
$GL_q(N)$- differential calculi.

Since all the type II differential complexes are isomorphic
(see first remark to the Theorem 1), we expect that $\sigma = -\kappa_q
[N]_q$ differential calculus (\ref{28}),(\ref{28a})
can be transformed to the case of any
$\sigma$. In order to realize this transformation we consider the new
set of generators $\{ T^{(g)}_{ij} \}$ of the algebra $Fun(GL_{q}(N))$
\be
\lb{37}
T^{(g)}_{ij} = g(z) T_{ij} \; ,
\ee
where $z = det_{q} (T)$ and $g(z)$ is an arbitrary function of $z$.
It is clear that $det_{q}(T^{(g)}) = zg(z)^{N}$, and therefore
the choice $g(z) = z^{-1/N}$ leads to the
$SL_{q}(N)$-case of \cite{SWZum,Zum}. Using commutation relations
(\ref{28}) and (\ref{30a}) one can deduce
\be
\lb{38}
z \, dT = q^{2} dT \, z \; , \;\; \lambda q^{N} dT = [T, \; \omega]
\Rightarrow \lambda q^{N} dg(z) = \o (g(q^{2}z) - g(z)) \; .
\ee
Now we introduce new Cartan 1-forms
$\Omega^{g} = dT^{(g)}(T^{(g)})^{-1}$
which relate to the old ones via the following formulae
\be
\lb{39}
\Omega^{g} = g dT T^{-1}g^{-1} + dg \cdot g^{-1} =  \Omega G(z) +
\o \frac{G(z) -1}{\lambda q^{N} } \; .
\ee
Here $G(z) = g(q^{2}z)g^{-1}(z)$.
Note, that eqs.(\ref{28}) and (\ref{30}) give the following
formulas
\be
\lb{40}
\lambda q^{N} d\Omega = - \{ \Omega , \o \} = \lambda q^{N} \Omega^{2}
\ee
Using these equations and relations (\ref{28}) and (\ref{30})
we obtain the set of
$GL_{q}(N)$-differential calculi parametrized by the function $G(z)$:
\be
\lb{41}
\begin{array}{c}
\R \T^{(g)} {\T^{(g)}}'  =  \T^{(g)} {\T^{(g)}}' \R \; , \nonumber \\ \\
\R \Omega^{g} \R \Omega^{g} +
\Omega^{g} \R \Omega^{g} \R^{-1}  =
\R U^{g} \R + U^{g} \; , \nonumber \\ \\
\T^{(g)} {\O^{g}} '  =  \R \O^{g} \R \T^{(g)} G(z)  +
\O^{g} \T^{(g)} (G(z)-1) +  \nonumber  \\
\o^{(g)} (1- \R^{2} G(z)) \T^{(g)}
\left( [N]_{q} + \frac{\lambda q^{N} G(z)}{G(z) -1} \right)^{-1}  \nonumber ,
\end{array}
\ee
where
\be
\lb{42}
U^{g}  =  (\Omega^{g})^{2} (1- G(z)) + \o \Omega^{g}
\frac{ G(q^{2}z) - G(z)}{\lambda q^{N}} \; ,
\ee
\be
\lb{43}
\o^{(g)} = Tr_{q} \Omega^{g} = \o \left( G(z) +
\frac{ [N]_{q}(G(z) -1)}{\lambda q^{N} } \right)
\ee

Let us consider the case when the function $G(z)$ is a constant.
For example for $g(z) = z^{\alpha}$ we have $G(z) = q^{2\alpha}$
and Eqs.(\ref{41}) give us the one-parametric set of differential calculi
which can be naturally correspoponded to the set of type II
differential complexes (\ref{116}):
\be
\lb{44}
\begin{array}{c}
\R \T^{(\alpha)} {\T^{(\alpha)}}'  =
\T^{(\alpha)} {\T^{(\alpha)}}' \R \; , \nonumber \\ \\
\R \O^{(\alpha)} \R \O^{(\alpha)} +
\O^{(\alpha)} \R \O^{(\alpha)} \R^{-1}  =
\mu_{\alpha} (\R (\O^{(\alpha)})^{2} \R +
(\O^{(\alpha)})^{2}) \; , \nonumber \\ \\
\T^{(\alpha)} {\O^{(\alpha)}} '  =
\R \O^{(\alpha)} \R \T^{(\alpha)} (1-\mu_{\alpha})  -
\O^{(\alpha)} \T^{(\alpha)} \mu_{\alpha} -  \nonumber \\
\left(  \frac{\mu_{\alpha}}{ \xi(\mu_{\alpha})} \right)
\R(1 + \frac{\mu_{\alpha}}{\lambda} \R )
\o^{(\alpha)} \T^{(\alpha)}
\nonumber ,
\end{array}
\ee
where $\mu_{\alpha} = 1-q^{2\alpha}, \; \xi(\mu_{\alpha}) =
q^{N}(1-\mu_{\alpha}) - \frac{\mu_{\alpha}}{\lambda}[N]_{q}$ and
\be
\lb{45}
\o^{(\alpha)} = Tr_{q} \Omega^{(\alpha)} =
q^{-N} \xi(\mu_{\alpha}) \o\; .
\ee

Now let us explore the possibilities of $SL_q(N)$-reduction
of this calculi. First, if we put
$\alpha = -1/N$ (as it was done in Refs.\cite{SWZum,Zum}),
then we have in the
commutation relations (\ref{44}) unavoidable additional
1-form generator $\o^{(-1/N)}$ and, thus, the number of
Cartan's 1-forms is $N^{2}$ but not $N^{2}-1$ as in the
undeformed case of $SL(N)$.
Second, one could try to put $\o^{(\alpha)}$ to zero choosing parameters
$\alpha$ and $q$ as
\be
\lb{46}
q^{-\alpha}\xi(\mu_{\alpha}) = q^{\alpha + N} + [\alpha]_{q} [N]_{q} = 0 \; .
\ee
In particular this equation is fulfilled for the $q$ being
a root of unity:
$ q^{-2N} = q^{\pm 2} = q^{2\alpha} $, which doesn't contradict
with the condition $\alpha = - 1/N$.
However, for the case of (\ref{46})
we have in third equation of (\ref{44}) $\;{0 \over 0}$-ambiguity,
which resolves as
\be
\lb{48}
q^{N - \alpha} \o^{(\alpha)}
\left(
q^{\alpha + N} + [\alpha]_{q} [N]_{q}
\right)^{-1}  = \o \; ,
\ee
and we can not put it to zero having in mind that
$\lambda q^{N} dT^{(\alpha)} = [ T^{(\alpha)}, \; \o ]$ and
$[det_{q} T^{(\alpha)}, \; \o ] \neq 0$. Therefore, the
differential calculi (\ref{44}) doesn't admit the correct
$SL_q(N)$-reduction even for the special values of the quantization
parameter $q$.

Now, how one may hopes to construct the consistent
bicovariant differential calculus on $SL_q(N)$?
The nice way of making the reduction from $GL_q(N)$-case
doesn't work for type II differential calculi (\ref{44}).
May be the cross-multiplication presented in (\ref{44}) is not
the unique possibility of construsting the differential calculi
starting from the type II complexes. May be the difficulties will be
overcomed if we use the type IA complexes instead of
the type II ones. But here for $\rho \neq 1$ we meet the serious
problems when constructing the local coordinate representation
of the type $\O = dT \cdot T^{-1}$.
So, only the type IA differential complex with $\rho = 1$
seems to be good candidate for construction of consistent
differential calculus on $GL_{q}(N)$ with it's possible reduction to
$SL_{q}(N)$.
We hope to revert to these problems
in further publications.

\section*{Acknowledegments}
We would like to thank A.T.Filippov, P.P.Kulish, F.M\"{u}ller-Hoissen and \\
A.A.Vladimirov for valuable disscussions.

\appendix
\section{Appendix}
\setcounter{equation}0

Here we present the 'ordered' expressions
for quadratic combinations $(\tO \circ \tO^2)_{21}$,
$(\tO^2~\circ~\tO)_{21}$ (see (\ref{i29})), and collect some formulas
which were used in derivation of these expressions.

Consider the sequence $x_i$ defined iteratively
\be
\lb{A1}
x_0 = q^N + q^{-N} \; , \;\;\; x_1 = [N]_q + \lambda q^N \; , \;\;\;
x_{i+2} = x_{i} + \lambda x_{i+1} \; .
\ee
Define
\be
\lb{A2}
y_i = (-1)^i {x_{1-i}x_1 - x_{-i}x_2 \over |x|} \; , \;\;\;
y_{i,k} = (-1)^{i+k} {x_{1-i-k}x_{1-k} - x_{-i-k}x_{2-k} \over |x|} \; ,
\ee
where $|x| \equiv [N+2]_q[N-2]_q$.
It is strightforward to show that $y_{i,k} = y_{i}$ for any $i$ and $k$, and
$y_i$ are calculated by the following
simple iteration:
\be
\lb{A3}
y_0 = 1 \; , \;\;\; y_1 = \lambda \; , \;\;\;
y_{i+2} = y_i + \lambda y_{i+1} \; .
\ee
When simplfying the final expressions for $(\tO \circ \tO^2)_{21}$ and
$(\tO^2 \circ \tO)_{21}$ we use the following properties of the matrix
functions $x(\R)$, $\bar{x}(\R)$:
$$
\R^k x(\R) = x_k{\bf 1} + x_{k+1}\R \; , \;\;\;
\R^k \bar{x}(\R) = {(-1)^k \over |x|}\left( -x_{2-k}{\bf 1}
 + x_{1-k}\R \right) \; ,
$$
$$
\R^k x_i \bar{x}(\R) = (-1)^{i+k}\left( y_{-i-k}{\bf 1} - x_{1-k}\R^{1-i}
\bar{x}(\R) \right) \; ,
$$
together with (\ref{A1})-(\ref{A3}). The result is
\be
\lb{A4}
\begin{array}{l}
\!\!\!\!\! (\tO^{2} \circ \tO)_{21} =
E^{-1}(\R) \left[ \epsilon(\R)\left\{
{\bf 1} + {x(\R) \over |x|}\delta(\R) \right\}
(\tO \circ \tO^{2})_{12}
\right. \\ \\
+ \R\left\{ \epsilon(\R) \left( \lambda -
{\delta x_{-1} + q^{N} x_{-3} \over |x|} \right) +
{\tau\sigma \over |x|}\left( [N]_{q} + \R\delta(\R) \right)
\right\} (\tO^{2} \circ \tO)_{12}
\\ \\
- \tau\R^{2}\epsilon(\R)\left\{ \bar{x}(\R) + \rho
{x(\R) \over |x|} \R^{-2} \right\} \o(\tO \circ \tO)_{12}
\\ \\
+ \left\{ {\delta x_{1} + q^{N}x_{-1} \over |x|} \R^{2}
({\bf 1} + \R^{2}\bar{x}(\R)) + {\tau\sigma \over |x|}\R \left(
[N]_{q} + \lambda q^{N}({\bf 1} + \R^{2}) + (1 + \lambda^{2})\R
\delta(\R) \right) \right\} (\tO^{3})_{1}
\\ \\
- \left\{ {\delta x_{1} + q^{N} x_{-1} \over |x|}\R
({\bf 1} + \R^{2}\bar{x}(\R))
+ { \tau\sigma \over |x|} \left( q^{N}({\bf 1} + \R^{2}) + \lambda
\R\delta(\R) \right) \right\} (\tO^{3})_{2}
\\ \\
+ \left\{ \tau\bar{x}(\R)\epsilon(\R)\left(
\lambda\R - { \delta x_{0} + q^{N} x_{-2} \over |x|} \right)
- {\tau\rho x_{3}  \over |x|}\R\epsilon(\R) +
\lambda q^{N}{\tau\rho \over |x|}\R^{-1} F(\R)
\right\} \o(\tO^{2})_{1}
\\ \\
\left.+ \left\{ - \tau\bar{x}(\R)\epsilon(\R) \left(
1 + {\delta x_{0} + q^{N} x_{-2} \over |x|} \right)
+ {\tau\rho x_{2} \over |x|}\epsilon(\R) +
\lambda q^{N} {\tau\rho \over |x|}\R^{-1} F(\R)
\right\} \o(\tO^{2})_{2} \right] \; ,
\end{array}
\vspace*{0.3cm}
\ee
\be
\lb{A5}
\begin{array}{l}
\!\!\!\!\!\! (\tO \circ \tO^2)_{21} =
E^{-1}(\R) \left[ \left\{ \R \epsilon(\R)
{\delta x_{1} + q^{N} x_{-1} \over |x|}
- {\tau\sigma \over |x|} \left(
\lambda q^N \R^3 - x_{-1}\R + \R^2\delta(\R) \right)
\right\}(\tO \circ \tO^2)_{12}
\right. \\ \\
\!\!\!\!\!\!\!\!
+ \left\{ \R^2\epsilon(\R)\left( {\bf 1} - \bar{x}(\R)\delta(\R)
+\tau\sigma{x_{-2} + x_0 \over |x|} \right) +
\tau\sigma(\R^4 + \R^6)\delta(\R)\bar{x}^2(\R)
- {(\tau\sigma)^2  \over |x|}\R^2 \right\}(\tO^2 \circ \tO)_{12}
\\ \\
- \tau\R^2 \left( \bar{x}(\R) + \rho {x(\R) \over |x|}
\R^{-2} \right) \biggl( \epsilon(\R) + \tau\sigma\R^2\bar{x}(\R)\biggr)
\o(\tO \circ \tO)_{12}
\\ \\
+ \left\{ {\delta x_1 + q^N x_{-1} \over |x|}\R^3 G(\R) + \tau\sigma
\R^2 \left({x_2 \over |x|} + {x_3 \over |x|}\R\bar{x}(\R)\delta(\R)
\right) - {(\tau\sigma)^2 \over |x|}(1 + \lambda^2)
\R^2 \right\} (\tO^3)_1
\\ \\
- \left\{ {\delta x_1 +q^N x_{-1} \over |x|} \R G(\R) + \tau\sigma\R\left(
{x_1 \over |x|} +{x_2 \over |x|}\R\bar{x}(\R)\delta(\R) \right)
- \lambda\R{(\tau\sigma)^2 \over |x|} \right\} (\tO^3)_2
\\ \\
+ \left\{ {\tau\rho \over |x|}\R^2 \biggl( \epsilon(\R)
\bigl( -x_2 + (1 + \lambda^2 - x_3 \bar{x}(\R)\R)\delta(\R) \bigr)
+ \tau\sigma(1+\lambda^2) \biggr) \right.
\\ \\
\left. + \tau \bar{x}(\R) \left( \epsilon(\R)\left(\tau\sigma{x_0 \over |x|}
+ \R^2 \right) - \lambda\tau\sigma\R^3\bar{x}(\R) \right)
\right\} \o(\tO^2)_1
\\ \\
+ \left\{ {\tau\rho \over |x|}\biggl( \R\epsilon(\R)\left(
x_1 + (x_2 \R \bar{x}(\R) - \lambda)\delta(\R) \right)
- \lambda\tau\sigma\R \biggr) \right.
\\
\left. + \tau^2\sigma\bar{x}(\R) \left( {x_0 \over |x|}\epsilon(\R)
+\R^2\bar{x}(\R) \right) \right\} \o (\tO^2)_2 \biggr] \; .
\end{array}
\ee
where
$$
\delta(\R) = \delta + q^N \R^2 \; , \;\;\;
\epsilon(\R) = {\bf 1} + \R^{2} \bar{x}(\R)\delta(\R)\; ,
$$
$$
F(\R) = {\bf 1} + \R^{-4} {x(\R) \over |x|} \delta(\R) \; , \;\;\;
G(\R) = \epsilon(\R) + \tau\sigma\R^2\bar{x}(\R) \; ,
$$
\be
\lb{A6}
E(\R) = \left( {\bf 1} +
{\delta x_{0} + q^{N}x_{-2} \over |x|} \right)\epsilon(\R)
+ {\tau\sigma \over |x|} \left( q^{-N} + q^{N}\R^{2} \right)\; .
\ee
Note, that these expressions are significantly simplified under
restriction $\tau\sigma = 0$.

\end{enumerate}
\end{document}